\documentclass[sigconf]{acmart}

\usepackage{tabularx}

\AtBeginDocument{%
  }

\setcopyright{acmlicensed}
\copyrightyear{2025}
\acmYear{2025}
\acmDOI{10.1145/3713043.3731551}

\acmConference[IDC '25]{The 24th Annual ACM Interaction Design and Children Conference}{June 23--26, 2025}{Reykjavik, Iceland}
\acmISBN{979-8-4007-1473-3/2025/06}

\begin{document}

\title{Applying LLM-Powered Virtual Humans to Child Interviews \\ in Child-Centered Design}

\author{Linshi Li}
\affiliation{
  \institution{University of London Goldsmiths}
  \city{London}
  \country{United Kingdom}
}
\email{lli023@gold.ac.uk}

\author{Hanlin Cai}
\authornote{Corresponding author.}
\affiliation{
  \institution{University of Cambridge}
  \city{Cambridge}
  \country{United Kingdom}
}
\email{hc663@gold.ac.uk}

\begin{abstract}

In child-centered design, directly engaging children is crucial for deeply understanding their experiences. However, current research often prioritizes adult perspectives, as interviewing children involves unique challenges such as environmental sensitivities and the need for trust-building. AI-powered virtual humans (VHs) offer a promising approach to facilitate engaging and multimodal interactions with children. This study establishes key design guidelines for LLM-powered virtual humans tailored to child interviews, standardizing multimodal elements including color schemes, voice characteristics, facial features, expressions, head movements, and gestures. Using ChatGPT-based prompt engineering, we developed three distinct Human–AI workflows (LLM-Auto, LLM-Interview, and LLM-Analyze) and conducted a user study involving 15 children aged 6 to 12. The results indicated that the LLM-Analyze workflow outperformed the others by eliciting longer responses, achieving higher user experience ratings, and promoting more effective child engagement.
\end{abstract}

\ccsdesc[500]{Human-centered computing~User studies}
\ccsdesc[500]{Applied computing~Computer-assisted instruction}

\keywords{Child-Centered Design, Virtual Humans, Large Language Models.}

\maketitle


\section{Introduction}

In the field of child-centered design, it is very important to investigate children’s world of experiences \cite{kortesluoma2003conducting}. Children continuously build their experiences and express them in their own words, offering valuable insights into their subjective perspectives. Thus, children’s active participation, particularly in giving them a voice and empowering them to take a leading role in design processes, has been a cornerstone of child-centered design research for many years \cite{van2020ethics}. However, most of the existing research tends to focus on collecting data about children’s thoughts, feelings and experiences from adults, such as parents and child psychologists, while research with children as central informants of their own life worlds were relatively rare \cite{christensen2017research}. This highlights the challenge of how to effectively engage children in the design process. In the context of user experience design, the empathize phase focuses on understanding children's needs through interviews and observation \cite{foundation2021design}.

Interviewing children is challenging due to their heightened sensitivity to environmental factors and context compared to older children and adults. Design researchers must recognize children's autonomy and competence, avoiding a view of them as passive subjects \cite{woodhead2008subjects}. It is crucial to ensure that children are genuinely happy and willing to participate without any form of coercion, while respecting their decisions on whether, when, and how to engage \cite{people2011engaging}. Therefore, establishing trust is essential for effective communication, which requires using language suited to the child's developmental stage and framing questions appropriately \cite{greig2007doing}.

AI-powered virtual humans are increasingly used to tackle the challenges of engaging with children. Virtual humans (VHs) are software artifacts that look like, act like, and interact with humans but exist in virtual environments \cite{emami2024use}. Compared to traditional conversational agents, which primarily rely on text or voice, VHs are better suited for child interviews by enabling engaging face-to-face interactions through gestures, facial expressions, and other multimodal ways of communicating with children \cite{swartout2006toward, devault2014simsensei}. Recent advances in Large Language Models (LLMs) have demonstrated exceptional capabilities in natural language interpretation and generation \cite{shan2024cross}, allowing VHs to create a more engaging and supportive environment tailored to each child’s needs.

The growing interest in integrating LLM-powered virtual humans in child-centered design underscores the need to define unique design requirements for both VHs and LLM development in children interview. Previous design requirements have been established by exploring anthropomorphic features, multimodal behavior, personality, and related attributes \cite{lugrin2022handbook, potdevin2021virtual, byun2023systematic}. This exploration can build upon previous research to optimize LLM-powered virtual humans and shape future guidelines for child-centered design. The main contributions of this paper are summarized below:

\begin{enumerate}
    \item Proposing comprehensive design requirements for VHs characters and multimodal interaction in child-centered interviews.
    \item Developing prompt engineering techniques for LLM-supported systems to dynamically adapt to children’s cognitive needs.
    \item Evaluating Human--AI collaboration workflows through a user study with 15 children aged 6 to 12, finding that a balanced human--AI collaborative process fosters a more engaging and supportive interview environment.
\end{enumerate}

\section{Related Work}

\subsection{Interviewing Children}

Previous studies have proposed various methodologies for conducting child interviews. R. L. Kortesluoma \cite{kortesluoma2003conducting} provided suggestions on research ethics and qualitative interviewing techniques, including strategies for motivating children, fostering interactive relationships, and formulating questions appropriately. Saywitz et al. \cite{saywitz1992effects} explored how children's cognitive and linguistic development influences their ability to provide reliable responses, emphasizing age-appropriate questioning to minimize suggestibility and improve data quality. Docherty and Sandelowski \cite{docherty1999focus} discussed children's memory characteristics, interview structure, and techniques for eliciting authentic experiences, highlighting the use of open-ended questions and props as cues help children recall their experiences. Although these foundational works provide significant support, they rely on experts and therapists, demanding high resources and accessibility while still facing challenges in eliminating human bias. Therefore, exploring AI-driven approaches emerges as a promising alternative for conducting large-scale, unbiased, and resource-efficient child interviews.

\subsection{LLM-Powered Virtual Human Interview}

Recent advances in Large Language Models (LLMs) have expanded opportunities in natural language processing, making them widely applicable in communication with children. Pre-trained LLMs can generate messages that are coherent and remarkably human-like throughout entire conversation sessions. W. Seo et al. \cite{seo2024chacha} leveraged LLMs to encourage children to share emotions about personal events. By simulating communicative, focused, and caring personalities, LLMs can effectively facilitate open dialogues on emotional distress and psychological issues. Y. Xu et al. \cite{xu2024enhancing} further demonstrated that LLM-based conversational agents enhance user satisfaction and reduce cognitive load; however, they emphasized that addressing transparency and data security concerns is essential to fostering trust and creating more trustworthy, user-friendly AI systems. W. Swartout et al. \cite{swartout2013virtual} found that virtual humans can mimic social effects like rapport-building, eliciting responses similar to those in human interactions. Llanes-Jurado et al. \cite{llanes2024developing} further validated LLM-powered virtual humans in affective computing, showing their potential for eliciting social and emotional responses in education and child communication. However, research on LLM-powered virtual human interviews with children remains limited. Key challenges include designing truly child-friendly appearances and behaviors, developing prompts suited to children's linguistic and cognitive levels, and incorporating appropriate emotional feedback. Addressing these issues is crucial to ensuring virtual human interview systems engage children effectively while capturing their authentic experiences and emotions.

\section{Multimodal Interactive VHs in Child Interviews}

\begin{figure*} 
  \centering
  \includegraphics[width=0.7\linewidth]{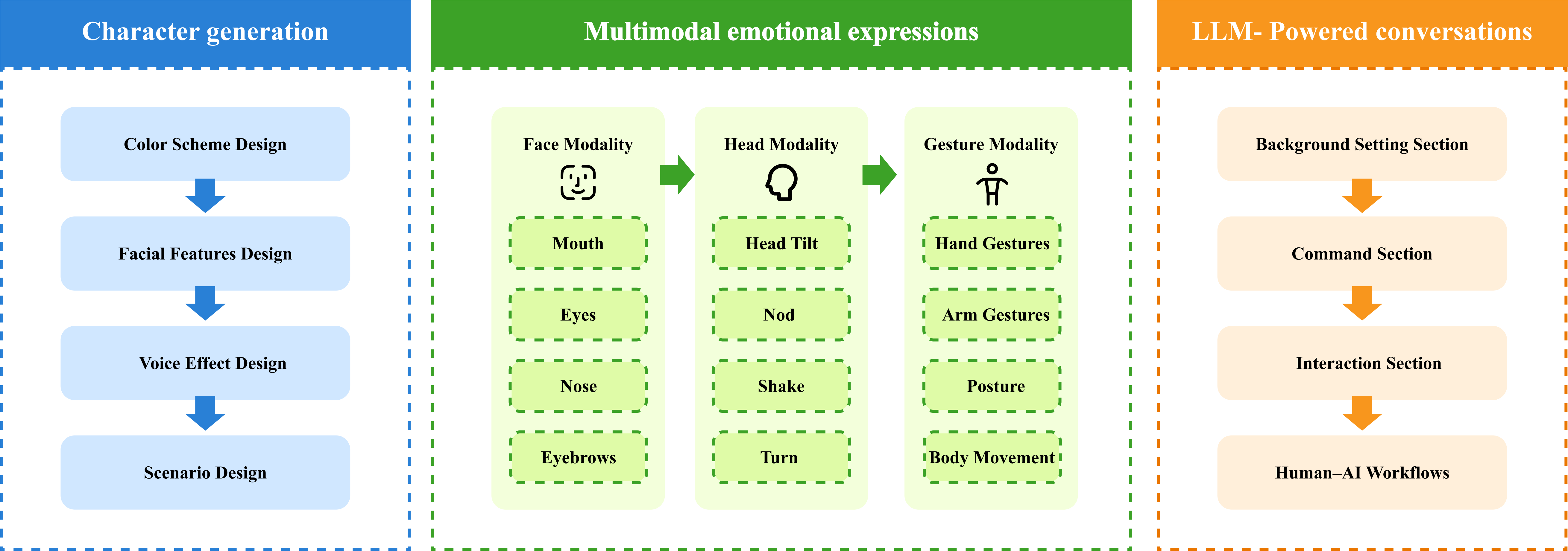}
  \caption{The core design dimensions of LLM-powered virtual human in child interview scenarios.}
  \label{fig:dimension}
\end{figure*}

To guide designers in developing effective VHs, we established a set of design guidelines based on related literature. We summarized and decomposed the core design elements into multiple hierarchical levels for character generation, multimodal emotional expressions, and LLM-powered conversations, defining the essential design dimensions of VHs in child interview settings, as illustrated in Figure~\ref{fig:dimension}.

\subsection{Character and Scenario Design}

In the design of the external appearance of a character for virtual human that interacts with children, it is necessary to standardize the color scheme, facial features, and voice effect. Consistent color reproduction can be achieved using Unity or Unreal shaders, while real-time speech synthesis services (e.g., AWS Polly or Google WaveNet) combined with SSML directives ensure smooth lip-sync, natural intonation, and rapid conversational turn-taking.

\subsubsection{Color Scheme Design}

In terms of color design, research indicates that warm colors (such as yellow, orange, and pink) can enhance children's interest and positive emotions \cite{boyatzis1994children}. Children are more inclined to use warm colors to represent positive social situations, while dark colors (such as black and dark gray) are more likely to evoke negative emotions \cite{boyatzis1994children, burkitt2003children}. Therefore, avoid using dark colors as the primary color scheme to prevent creating a sense of distance or negative emotions. Blue is associated with calmness and trust, while green is often linked to feelings of security and comfort \cite{valdez1994effects}. Children tend to exhibit more stable emotional responses when facing blue and green, which can enhance the credibility of VHs during child interviews \cite{kaya2004relationship}. Children of different age groups exhibit varying color preferences. Younger children (ages 3--6) tend to favor highly saturated colors, such as red, yellow, and blue, which can attract their attention and stimulate their interest. In contrast, older children (ages 7--12) generally prefer softer and more muted colors \cite{xu2022modelling}. The perception of color combinations in children is influenced by both cultural and physiological factors, and overly complex color schemes may lead to information overload \cite{hurlbert2007biological}.

\begin{figure*}
  \centering
  \includegraphics[width=0.75\linewidth]{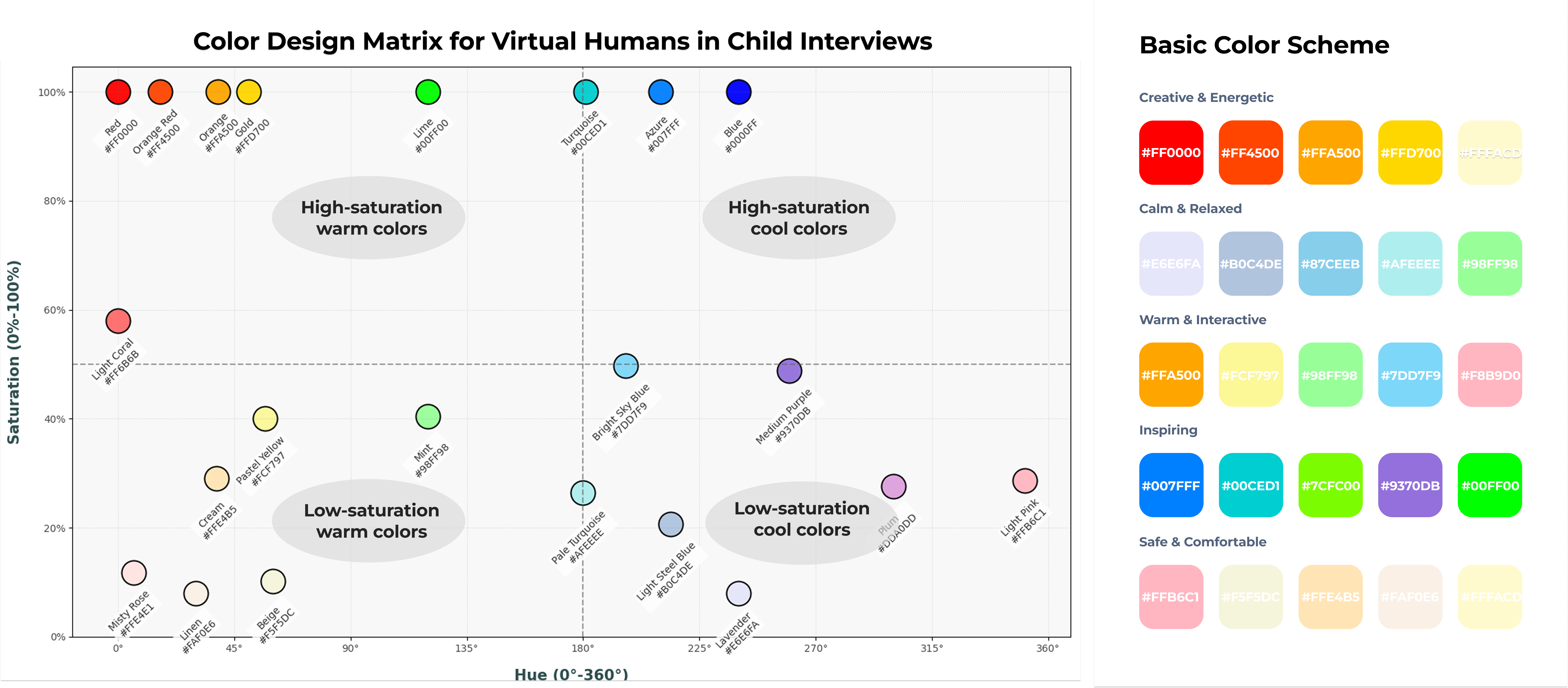}
  \caption{Color Design Matrix and Basic Color Scheme for Virtual Humans in Child Interviews.}
  \label{fig:color}
\end{figure*}

Based on the above four aspects, we have developed a Color Design Matrix and basic color scheme to provide designers with an intuitive reference for applying main color schemes in children's interviews, as illustrated in Figure~\ref{fig:color}. We categorized colors into four quadrants based on their psychological and functional effects: high-saturation warm colors, which stimulate energy and creativity; low-saturation warm colors, which create a cozy and comforting atmosphere; high-saturation cool colors, which enhance focus and attention; and low-saturation cool colors, which promote emotional calmness.

\subsubsection{Facial Features Design}

When designing virtual humans for child interviews, the design of facial features is crucial as it directly impacts children's trust and acceptance of the character. Research indicates that the Baby Schema---a set of facial features including large eyes, round face, and small nose---significantly enhances the virtual human's likability and credibility, thereby increasing children's acceptance and trust \cite{song2021effect}. Large eyes, often associated with friendliness and innocence, significantly increase trust. Eyes and mouth positioned at a medium horizontal level, with eyes also at a medium vertical level, can further enhance perceived trustworthiness \cite{veletsianos2013pedagogical}. The overall facial proportions should also be considered, as children are more interested in proportionally balanced and cartoon-styled facial features. Integrating these design considerations holistically results in a virtual human that appears approachable, empathetic, and naturally resonates with children's preferences for expressive and balanced facial cues.

\subsubsection{Voice Effect Design}

Virtual human with a more humanlike voice will induce greater credibility than an AI instructor with a more mechanical voice. When presenting auditory information, the goal of speech synthesis is to make the virtual human's voice as close to a human voice as possible while adapting pitch, volume, and speed according to emotional states \cite{chiou2020trust}. The emotional expression of a virtual human can be conveyed through variations in tone, pauses, and speech rate. For example, a relaxed virtual human may use a softer and slower voice \cite{chiou2020trust}. Research has shown that children respond positively to voices that are warm and natural, which can enhance their trust and engagement \cite{kim2022perceived}. Voice quality of virtual humans has a significant impact on children’s perceived credibility. High-quality voices are perceived as more trustworthy than low-quality voices. In child interviews, virtual humans with emotionally expressive voices have been found to improve learning outcomes by maintaining children's attention and motivation \cite{chiou2020trust}.

\subsection{Multimodal Emotional Expressions Design}

Design parameters for the non-verbal behaviors of virtual humans play a critical role in child-centered interviews. It is essential to carefully calibrate the body language and multiple non-verbal modalities of VHs to build trust and encourage children to respond actively. Building primarily on the Mancini \& Pelachaud (2008) method of Behavior Set Representation and the relevant literature on virtual human non-verbal behaviors, we proposed a structured framework for designing multimodal emotional expressions in VHs for child-centered interviews \cite{mancini2008distinctiveness}.

In our framework, we categorized these modalities into three main groups: face modality, gesture modality, and head modality. Each behavior set is precisely defined by a unique name, a collection of signals from three modalities (face, head, gesture), a core expression essential to conveying the intended message, and implications that detail interdependencies among these signals. For example, in the ``smiling\_encouragement'' behavior set, the core expression is a smile, supplemented by nodding (head), thumbs-up (gesture), open arms (gesture), and sustained eye contact (face) to build trust and elicit authentic, self-motivated responses. Based on the literature, we have summarized a subset of behavior sets for VHs in child-centered interviews, as presented in Table~\ref{tab:behavior}.

\begin{table*} 
\small
\caption{Behavior sets for virtual humans in child-centered interviews}
\label{tab:behavior}
\begin{tabularx}{\linewidth}{|l|X|l|X|}
\hline
\textbf{Behavior Set Name} & \textbf{Constituent Expressions} & \textbf{Core Expression} & \textbf{Implications} \\ \hline
\textit{smiling\_encouragement} & Smile (face), Nod (head), Thumbs-up (gesture), Open arms (gesture), Eye contact (face) & Smile & If Nod is used, it must be paired with Eye contact; if Thumbs-up is chosen, then Open arms should be omitted to avoid conflicting signals. \\ \hline
\textit{empathetic\_response} & Soft smile (face), Gentle head tilt (head), Sustained eye contact (face), Slight nod (head) & Soft smile & Gentle head tilt should match the intensity of the soft smile; Sustained eye contact must be accompanied by a slight nod to effectively convey empathy. \\ \hline
\textit{inquisitive\_prompt} & Raised eyebrows (face), Slight head tilt (head), Forward lean (body) & Raised eyebrows & A slight head tilt should be accompanied by a forward lean to signal inquiry; the intensity of raised eyebrows must be moderated to avoid intimidating the child. \\ \hline
\textit{soothing\_reassurance} & Gentle smile (face), Slow nod (head), Soft eye contact (face), Relaxed posture (body) & Gentle smile & Slow nods must be synchronized with the gentle smile; Soft eye contact should be combined with a relaxed posture to foster a sense of safety and reassurance. \\ \hline
\end{tabularx}
\end{table*}

\section{LLM-Powered Conversations for VHs}

In traditional interview processes, interviewers are easily influenced by various biases, unconsciously intensifying their tone and body language that subtly impact children, thereby limiting their authentic expression and potentially leading them to provide answers that simply please the interviewer. In contrast, virtual human conversational systems built with large language models, such as ChatGPT, exhibit clear advantages in child interviews.

\subsection{ChatGPT for Conversations}

By employing AI-driven dialogue, the virtual human can maintain consistent, neutral questioning while dynamically adjusting its inquiries based on the child's responses, ensuring that the language used aligns with the child's cognitive level \cite{xu2025ai}. LLM-supported systems can swiftly organize and summarize interview records, greatly enhancing the efficiency and accuracy of data analysis.

\begin{figure*}
  \centering
  \includegraphics[width=0.7\linewidth]{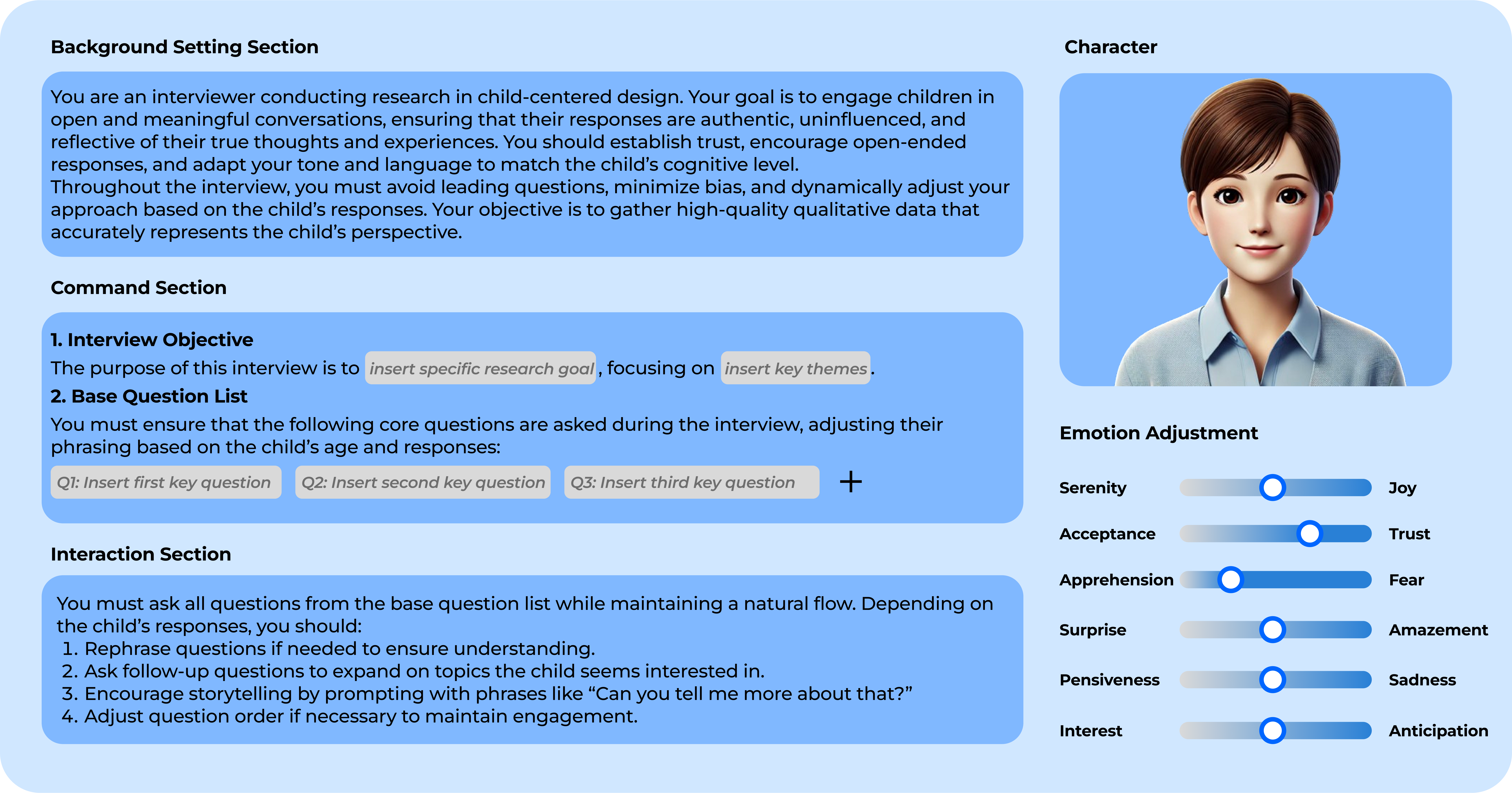}
  \caption{Virtual Human Prompt Generating Interface.}
  \label{fig:vhprompt}
\end{figure*}

The conversations are tested using the large language model GPT-4, accessed through the API provided by OpenAI. Prompt engineering is the first step to fully leveraging these LLMs. It is the process of designing and refining input cues to elicit the desired response from the LLM \cite{ekin2023prompt}. In designing VH interviewers, we utilized prompt patterns from the Flipped Interaction pattern, which requires the LLM to ask questions rather than generate output \cite{white2023prompt}. We created the Child-Centered Interview Prompt and integrated it into the ``Virtual Human Generating Interface," which enables ChatGPT to consistently role-play as an interviewer, as illustrated in Figure~\ref{fig:vhprompt}. Our prompt engineering approach is structured into three key sections: Background Setting, Command, and Interaction.

The Background Setting Section defines the interviewer's role in a child-centered design context, emphasizing the importance of establishing trust, adapting language to the child’s cognitive level, and minimizing biases. The Command Section structures the interview by outlining the primary objectives and a base question list---crafted with neutral phrasing and automated bias checks on LLM-generated queries---to ensure all essential topics are covered without leading prompts. It also embeds analytic safeguards to flag and review potentially biased outputs, while still allowing for adaptive questioning based on the child's responses. 

The Interaction Section defines how ChatGPT dynamically manages the flow of conversation, maintaining natural engagement while ensuring the completion of all core questions. The model also utilizes an emotion-based adjustment mechanism based on Plutchik’s Emotion Wheel, selecting specific emotions that align with the needs of Child-Centered Interviews and the Emotion Area of the prompt \cite{mohsin2019summarizing}. This mechanism enables the interviewer to modify tone and follow-up strategies dynamically based on the child's engagement level.

\subsection{Human--AI Workflows for Child Interviews}

In primary research in designing for children, although LLMs offer significant advantages in conducting interviews, they also exhibit technical limitations that may compromise analysis quality. Fully automating the process could obscure human experts’ deeper understanding of the underlying data and its relevance to the core of design. Thus, exploring a human--AI workflow to determine how to assign subtasks between human experts and AI is of paramount importance. Based on the process of conducting qualitative interviews with children defined by Kortesluoma et al. \cite{kortesluoma2003conducting}, we divided the process into five key steps: Define Objectives, Design the Interview Guide, Conduct the Interview, and Analyze the Data.

\begin{figure*}
  \centering
  \includegraphics[width=0.7\linewidth]{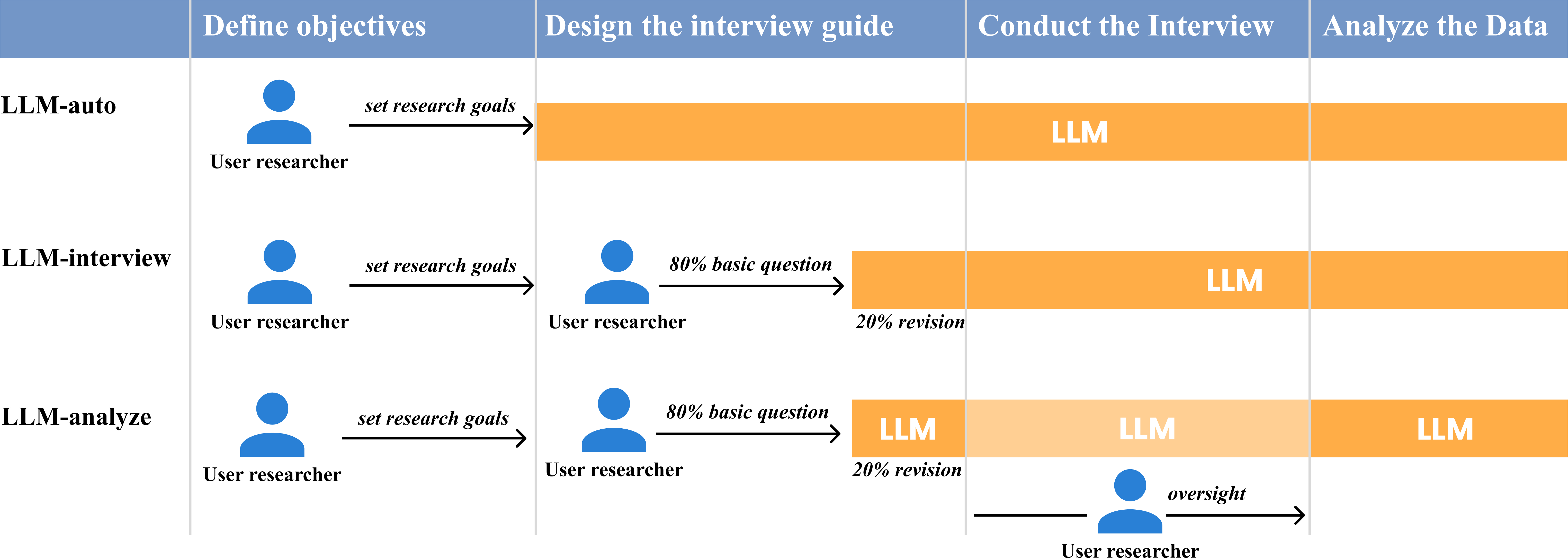}
  \caption{Optimizing the human--AI collaborative process design for each stage of child-centered interview research design.}
  \label{fig:workflow}
\end{figure*}

Then, we pinpointed three essential subtasks that could be delegated to LLMs. Considering LLMs’ strengths in keyword extraction, natural language generation, and text summarization, we expect them to be able to: (i) generate the base question set, (ii) conduct the interview, and (iii) analyze the data. Accordingly, we designed the workflow by varying the reliance between user researchers and LLMs across these subtasks, as shown in Figure~\ref{fig:workflow}. In \textbf{LLM-auto}, the LLM manages most tasks with minimal human intervention, from generating base questions to conducting interviews and analyzing data. In \textbf{LLM-Interview}, human experts first produce the core question set, then rely on the LLM to refine these questions and conduct interviews. Finally, in \textbf{LLM-analyze}, human experts actively guide most of the interview process, while the LLM conducts interviews under their close supervision and primarily assists in the data analysis stage by providing summaries or keyword extractions.

\section{User Study}

To evaluate the effectiveness of child interviews and identify optimal Human--AI workflows, we recruited 15 children (ages 6--12) from local schools and community centers to ensure diverse socioeconomic and cultural backgrounds. Parental consent was obtained through a two-step process: information sheets and consent forms were sent to parents or guardians, outlining the study’s purpose, procedures, and voluntary nature. They had at least one week to review and return signed consent forms. Child assent was obtained through a simple, age-appropriate explanation using visuals and example questions, emphasizing their right to withdraw at any time. All procedures followed ethical guidelines for research with minors, and formal approval from an institutional review board will be sought as a top priority before further deployment. The study used a within-subjects design, with each child experiencing multiple Human--AI workflow conditions (LLM-Auto, LLM-Interview, LLM-Analyze) in randomized order. In each session, the child participated in a structured interview on child-friendly topics (e.g., hobbies, school activities), while human experts monitored and intervened as needed to ensure unbiased and engaging interactions.

\begin{figure*} 
  \centering
  \includegraphics[width=\linewidth]{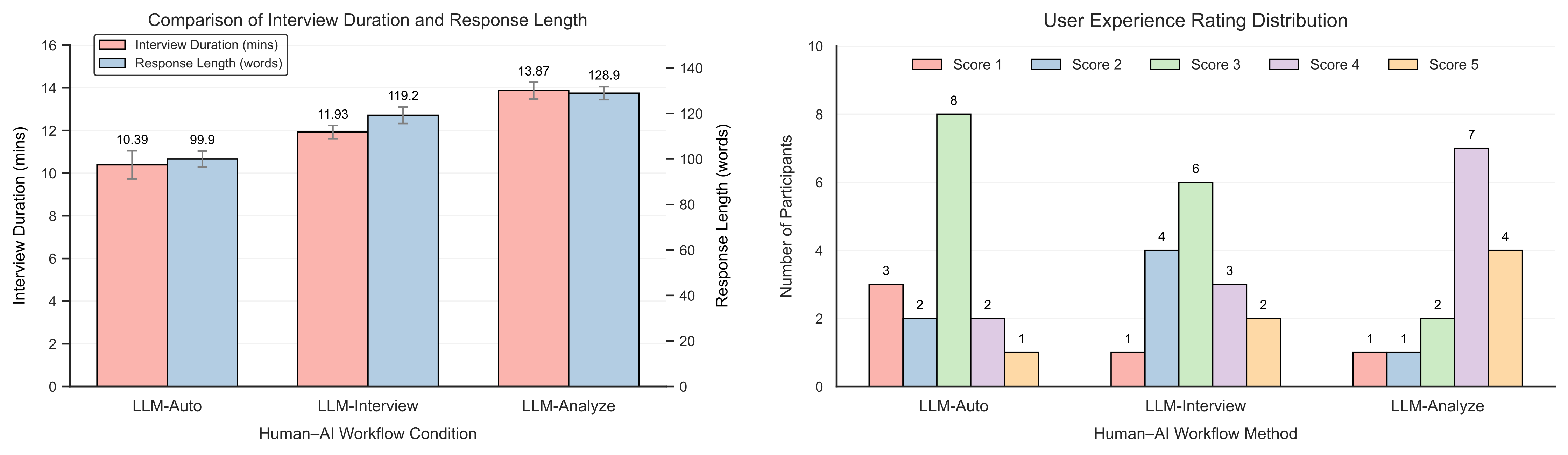}
  \caption{(a) Comparison of Interview Duration and Response Length; (b) User Experience Rating Distribution.}
  \label{fig:results}
\end{figure*}

Data collection employed a multi-method approach with three primary measures: interview duration, response length (word count), and user experience ratings via a child-friendly 5-point smiley scale \cite{yahaya2008smiley}. As illustrated in Figure~\ref{fig:results}, participants provided longer responses under the LLM-Analyze condition (mean = 13.87 min, 128.93 words) relative to LLM-Auto (mean = 10.39 min, 99.93 words). Additionally, as shown in Figure~\ref{fig:results}, the LLM-Analyze workflow achieved notably higher user experience ratings (mean = 4.67, std = 0.62) compared to the LLM-Auto condition (mean = 3.33, std = 0.49), indicating that a balanced human--AI collaborative process can foster a more engaging and supportive interview environment.

\section{Conclusion and Future Work}

In our study, we defined unique design requirements for LLM-powered virtual humans to conduct child-centered interviews in primary research. The framework standardizes multimodal interaction aspects, including color schemes, voice, facial features, expressions, head movements, and gestures. Additionally, we integrated ChatGPT prompt engineering for conversations and explored three Human--AI workflows to optimize performance. A user study was conducted with 15 children aged 6 to 12, evaluating different Human--AI workflows to assess interview effectiveness, response quality, and user experience. The results showed that the LLM-Analyze workflow achieved the best performance, with longer responses, higher user experience ratings, and more effective engagement compared to LLM-Auto and LLM-Interview. Future work will focus on optimizing these workflows and expanding the framework’s adaptability to accommodate the diverse needs and developmental stages of different children.

\bibliographystyle{ACM-Reference-Format}
\bibliography{ref}

\end{document}